\documentclass{article}

\usepackage{amssymb, amsmath}

\begin{document}

\title{On the Majorana equation -- Relations between its complex two-component and
real four-component eigenfunctions \newline }

\author{ {\bf Eckart Marsch} }

\date{}

\maketitle

{\bf }

Department for Extraterrestrial Physics, Institute for Experimental and
Applied Physics, Christian Albrechts University at Kiel, Leibnizstr. 11,
24118 Kiel,
Germany \\\\\\\\\\

{\bf Abstract  \newline}

We first derive without recourse to the Dirac equation the two-component
Majorana equation with a mass term by a direct linearization of the
relativistic dispersion relation of a massive particle. Thereby, we make only
use of the complex conjugation operator and the Pauli spin matrices,
corresponding to the irreducible representation of the Lorentz group. Then we
derive the complex two-component eigenfunctions of the Majorana equation and
the related quantum fields in a concise way, by exploiting the so called
chirality conjugation operator that involves the spin-flip operator.
Subsequently, the four-component spinor solutions of the real Majorana
equation are derived, and their intrinsic relations with the spinors of the
complex two-component version of the Majorana equation are revealed and
discussed extensively.

\newpage

\section{Introduction}
\label{s1}

In this paper we first present a new derivation \cite{marsch} of the
two-component Majorana equation including a mass term by a direct
linearization of the relativistic dispersion relation of a massive particle,
in a way similar to that used originally by Dirac \cite{dirac}. We
subsequently derive from the two-component complex the four-component real
version of the Majorana equation \cite{majorana}. We thus obtain some new
results and look at the problem from a fresh theoretical perspective, thus
expanding the established theoretical framework as described in modern
monographs \cite{mohapatra,fukugita} on this subject. One intention of this
work is to obtain the Majorana equation ``completely on its own rather than
as an afterthought when treating the Dirac equation", as it was phrased by
Case \cite{case}, who first reformulated half a century ago the Majorana
theory of the neutrino. Another main objective is to address the key question
of whether Majorana particles are their own antiparticles, which they are
according to common textbook wisdom. We will show that a real four-component
or complex two-component spinor wavefunction does not imply this conclusion
necessarily, but following Mannheim \cite{mannheim} we shall consistently
define antiparticles as being associated with the negative root of the
dispersion relation and particles with the positive one.

The recent paper by Pal \cite{pal} titled ``Dirac, Majorana, and Weyl
fermions" discussed the definitions of and connections between these three
fields, and their intrinsic symmetries. We consider the present paper partly
as a complement to his largely tutorial work, yet here with emphasis on the
two-component Majorana equation for a massive fermion, which is treated
without recourse to the Dirac equation \cite{marsch}. Thereby we only make
use of the complex conjugation operator and the Pauli spin matrices,
corresponding to the irreducible representation of the Lorentz group. The
Lorentz-invariant complex conjugation \cite{pal} involves the here introduced
spin-flip operator. Its connection to chiral symmetry is also discussed.
Finally, we also show that Dirac's equation in the real Majorana
representation is a direct consequence of the two-component complex Majorona
equation as obtained below.

We derive in much detail the eigenfunctions of the Majorana equation and the
related quantum fields, yet other than in the paper by Case \cite{case} in a
direct way, whereby we will exploit the spin-flip operator, or as we also may
call it perhaps more adequately, the chirality conjugation operator. Pal
\cite{pal} gave in the introduction to his paper a good motivation, which we
fully agree with, for dealing again with this theoretical subject. Our main
motivation is to contribute to the ongoing discussion of whether massive
neutrinos are Dirac or Majorana fermions, and to better understand the latter
theoretically in terms of the two-component theory and its connection to the
four-component real version. In this respect we also complement the old
pedagogical review on Majorana masses by Mannheim \cite{mannheim}.

\section{New derivation of the Majorana equation and its symmetries}
\label{s2}

In this section a new derivation of the two-component Majorana equation is
presented, by linearizing the standard dispersion relation of a massive
relativistic particle. Following Dirac \cite{dirac} historically, one can
derive his equation in a straightforward way from the relativistic energy and
momentum relation, $E = \sqrt{ m^2 + \mathbf{p}^2}$, which in fact is
constitutive for all field theories of massive particles. Here the
four-momentum of the particle is denoted as $p^\mu=(E,\mathbf{p}$), and the
units are such that the speed of light $c=1$, and the particle mass is $m$.
The contravariant four-momentum operator in space-time is denoted by $P^\mu=
(P_0, \mathbf{P}$), and its covariant form is given by the differential
operator $P_\mu=(P_0, -\mathbf{P}) = \mathrm{i}\partial_\mu =
\mathrm{i}(\partial/\partial t,\partial/\partial \mathbf{x})$ (conventionally
we use units of $\hbar=1$). The energy-momentum relation is usually written
in manifestly covariant form as mass-shell condition, $p^\mu p_\mu = m^2$.
Dirac's equation results from a its linearization, which is achieved by
introducing four new operators (usually represented by matrices), namely the
matrix three-vector $\boldsymbol{\alpha}$ and scalar $\beta$, yielding a
Hamiltonian in the linear algebraic form
\begin{equation}
\label{eq:2}
\mathcal{H} = \beta \,m + \boldsymbol{\alpha} \cdot \mathbf{\mathbf{P}},
\end{equation}
As shown in any text book (see, e.g., \cite{das,kaku}), in terms of matrices
one requires at least four dimensions to represent the operators
$\boldsymbol{\alpha}$ and $\beta$. Furthermore, to satisfy the mass-shell
requirement they must algebraically obey the (indicated by the symbol
$\{,\}$) anti-commutator,
\begin{equation}
\label{eq:3}
\{\boldsymbol{\alpha}, \beta \} = \mathbf{0},
\end{equation}
and also have the property that $\beta^2=1$ and
$(\boldsymbol{\alpha}\cdot\mathbf{p})^2=\mathbf{p}^2$. This is readily
ensured, for example in the chiral representation, if we decompose the
$4\times4$ matrix as follows, $\boldsymbol{\alpha}=\bar{\alpha}
\boldsymbol{\sigma}$, with $\bar{\alpha}^2=1$. Thus, $\boldsymbol{\alpha}$
can concisely be expressed in terms of the spin operator
$\boldsymbol{\sigma}$ and $\bar{\alpha}$. The $2\times2$ Pauli matrices are
known to have the required property
$(\boldsymbol{\sigma}\cdot\mathbf{p})^2=\mathbf{p}^2$. Similarly, we can
write $\beta= \bar{\beta} \sigma_0$. Finally, we quote explicitly the two
real matrices $\bar{\alpha}$ and $\bar{\beta}$ (and a related one called
$\bar{\gamma}= \bar{\beta} \bar{\alpha}$), which are going to be used later
and are defined as
\begin{equation}
\label{eq:4} \bar{\alpha} = \left(
\begin{array}{cc}
1  &  0  \\
0  & -1  \\
\end{array}
\right), \;\;\; \bar{\beta}= \left(
\begin{array}{cc}
0  &  1 \\
1  &  0 \\
\end{array}
\right), \;\;\; \bar{\gamma}= \left(
\begin{array}{cc}
0  &  -1  \\
1  &   0  \\
\end{array}
\right).
\end{equation}
The three Pauli matrices \cite{pauli}, to which we add here the $2\times2$
unit matrix denoted as $\sigma_0$, have their standard form, but are quoted
again because of their importance to what follows. They read
\begin{equation}
\label{eq:5} \sigma_{\rm x} = \left(
\begin{array}{cc}
0  &  1  \\
1  &  0  \\
\end{array}
\right), \;\;\; \sigma_{\rm y} = \left(
\begin{array}{cc}
0  &  -i  \\
i  &   0  \\
\end{array}
\right) , \;\;\; \sigma_{\rm z}= \left(
\begin{array}{cc}
1  &   0  \\
0  &  -1  \\
\end{array}
\right),
\end{equation}
and build the three-vector $\boldsymbol{\sigma}=(\sigma_{\rm x},\sigma_{\rm
y},\sigma_{\rm z})$. We may also introduce the two four-vector forms of the
Pauli matrices. They are defined according to Jehle \cite{jehle} by the
four-vectors $\sigma^\mu_\pm = (\sigma_0, \pm \boldsymbol{\sigma})$ and obey
$\sigma^\mu_\pm = \sigma_{\mp \mu}$. Making use of the above matrices, we can
obtain Dirac's equation in standard chiral form \cite{marsch, kaku} and the
classical Hamiltonian representation.

Returning to the basic equation (\ref{eq:2}), the question may then be asked
whether there are other than pure matrix representations of the operators
$\boldsymbol{\alpha}$ and $\beta$. In a recent paper \cite{marsch} it was
shown that the answer is yes. As this subject is not common knowledge, we
shall repeat here some of the key algebra, making use of the operator of
complex conjugation $\mathbb{C}$, which transforms any complex number $z$
into its complex conjugate $z^*$. This appears naturally in the symmetry
operations of time inversion and charge conjugation of Dirac's equation
\cite{kaku}. Therefore let us define an important operator, which is not of
pure algebraic nature but involves this complex-conjugation operator
$\mathbb{C}$. We call (in partial nomenclature analogy to $\beta$) this
operator $\tau$ and define it appropriately as $\tau=\sigma_\mathrm{y}
\mathbb{C}$, which differs from the definition in Marsch's paper
\cite{marsch} by an unimportant phase factor $\mathrm{i}$. The importance of
$\tau$ was first recognized by Case \cite{case}. It is anti-hermitian
conjugate, $\tau^\dagger = -\tau $, and other than $\beta$ obeys the relation
$\tau^2=-1$. Obviously, the operation of $\tau$ on the spin vector
$\boldsymbol{\sigma}$ leads to its inversion, i.e., when using the
commutation rules of the spin operators, we can show that the operation $\tau
\boldsymbol{\sigma} \tau^{-1} = -\boldsymbol{\sigma}$ yields a spin flip.
Also remember from equation (\ref{eq:5}) that $\boldsymbol{\sigma}^* =
\boldsymbol{\sigma}^\mathrm{T}$, where the superscript T indicates the
transposed matrix. As $\tau$ flips the spin, we have $\tau \sigma^\mu_\pm
=\sigma^\mu_\mp \tau $, and $\tau\mathrm{i}+\mathrm{i}\tau=0$, because of the
action of $\mathbb{C}$. Therefore, $\tau$ also anti-commutes with the
momentum four-vector operator $P_\mu$, and thus we have $\tau P_0 + P_0\tau
=0 $ and $\tau\mathbf{P} + \mathbf{P}\tau =0$.

Making use of the above properties of the spin flip operator $\tau=
\sigma_\mathrm{y} \mathbb{C}$, we can now still use a two-component matrix
representation given by the Pauli matrices (\ref{eq:5}). Therefore, by help
of $\tau$ we can go a decisive mathematical step beyond pure matrix algebra
and define the linear energy operator
\begin{equation}
\label{eq:7} \mathcal{H}= \tau \, m  + \boldsymbol{\sigma} \cdot \mathbf{P},
\end{equation}
in analogy to the equation (\ref{eq:2}). The operators involved obey the
anti-commutation rule,
\begin{equation}
\label{eq:8}
\{\boldsymbol{\sigma}, \tau \} = \mathbf{0},
\end{equation}
and also have the property that
$(\boldsymbol{\sigma}\cdot\mathbf{p})^2=\mathbf{p}^2$, but $\tau^2=-1$. We
stress again that for the space-time operators $\{ \mathbf{P}, \tau \} =
\mathbf{0}$ and $\{ P_0,\tau \} = 0$. When putting $P_0= \mathcal{H}$ in the
equation (\ref{eq:7}), squaring it and multiplying out, the above key
features of $\tau$ must be exploited. Thus we retain, when inserting the
differential operators explicitly, the Klein-Gordon equation.

As a consequence of (\ref{eq:7}), and without recourse to the Dirac equation,
we obtain directly a linear wave equation from (\ref{eq:7}), which is named
after Weyl \cite{weyl} without the mass term and after Majorana with the mass
term. It involves only the Pauli matrix operators acting on a two-component
spinor $\phi$, but introduces the complications that are caused by the
operator $\tau$, and reads as follows:
\begin{equation}
\label{eq:10} \mathrm{i}\left( \frac{\partial}{\partial t} +
\boldsymbol{\sigma} \cdot \frac{\partial}{\partial \mathbf{x}} \right)
\phi(\mathbf{x},t) = m \tau \phi(\mathbf{x},t).
\end{equation}
In (\ref{eq:10}) we have not indicated the $2\times2$-unit matrix explicitly.
This equation is nothing but what is called nowadays the two-component
Majorana equation \cite{mohapatra,fukugita}, although it was never written
down this way by Majorana himself. As shown by Case \cite{case}, this
equation can be also derived from Dirac's equation in chiral form, or more
generally if one imposes on it the condition of Lorentz-covariant complex
conjugation, a procedure which is clearly described in a recent tutorial
paper of Pal \cite{pal}. However, the view we take here is that equally well
we may consider equation (\ref{eq:7}) as basic, in the spirit of the citation
\cite{case} quoted in the introduction.

Obviously, there is a second version of the Majorana equation (\ref{eq:10}),
which is obtained by the operation of $\tau$ on it. Namely, when we apply
$\tau$ from the left side we find that
\begin{equation}
\label{eq:11} \mathrm{i}\left( \frac{\partial}{\partial t} -
\boldsymbol{\sigma} \cdot \frac{\partial}{\partial \mathbf{x}} \right)
\tau\phi(\mathbf{x},t) = - m \tau (\tau \phi(\mathbf{x},t)) = m \phi(\mathbf{x},t).
\end{equation}
Consequently, if $\phi$ solves the equation (\ref{eq:10}) with the plus sign
in front of $\boldsymbol{\sigma}$, then $\chi = \tau \phi$ solves it with the
minus sign, but with a minus sign also at the mass term. Effectively, this
amounts to replacing $\boldsymbol{\sigma}$ by $-\boldsymbol{\sigma}$ in
(\ref{eq:10}), also implying $\tau$ is exchanged by $-\tau$.

Apparently, the two equations (\ref{eq:10}) and (\ref{eq:11}) are closely
connected twins. It is interesting to note that the two together can provide
a special but important four-component Dirac spinor in the form
\begin{equation}
\label{eq:12} \psi^\mathcal{C}_\pm = \left(
\begin{array}{c}
\phi  \\
\pm \tau\phi \\
\end{array}
\right),
\end{equation}
which solves the Dirac equation. Moreover, charge (or better chirality)
conjugation \cite{kaku,das} is, in the chiral representation, given by the
operator $\mathcal{C}=\bar{\gamma}\tau$, which when operating on
$\psi^\mathcal{C}_\pm$ reproduces it. Since $\mathcal{C}^2 = 1$, this
operator has only the eigenvalue $\pm1$. Therefore, the spinors
$\psi^\mathcal{C}_\pm$ constructed from the solution of the Majorana equation
are the two eigenfunctions with eigenvalue $\pm1$ of the chirality
conjugation operator $\mathcal{C}$. Moreover, the chiral projection operators
are given by $P_\pm=1/2(1\pm \bar{\alpha})$, and thus the left- and
right-chiral spinor is $\psi_\mathrm{R,L}= P_\pm \psi$. As $\{ \bar{\gamma},
\bar{\alpha} \} =0$, we obtain $\mathcal{C} P_\pm = P_\mp \mathcal{C}$, and
consequently, if one applies $\mathcal{C}$ separately to the right- or
left-chiral fields, one obtains
\begin{equation}
\label{eq:13} \mathcal{C} [(\psi^\mathcal{C})_\mathrm{R}] =
(\psi^\mathcal{C})_\mathrm{L} \;\; \mathrm{and} \;\; \mathcal{C}
[(\psi^\mathcal{C})_\mathrm{L}] = (\psi^\mathcal{C})_\mathrm{R}.
\end{equation}
Thus this operator exchanges the chirality of the chiral projections of the
complex-self-conjugate spinor $\psi^\mathcal{C}$, and it transforms left-
into right-handed states and vice versa, a virtue which is essentially
associated with the spin-flip operator $\tau$ as key ingredient of
$\mathcal{C}$.

In conclusion, in this special situation the operator $\mathcal{C}$ acts as
chirality conjugation or reversal. While being a composite of the two related
and by themselves irreducible Majorana equations, the reducible chiral Dirac
equation has as an intrinsic symmetry this chirality conjugation symmetry.

Consequently, in what follows we can interpret the spinor $\phi$ as
representing a right-chiral field, and vice versa $\chi=\tau\phi$ as
representing a left-chiral field, which is obtained from $\phi$ by
application of the spin-flip operator. Apparently, it translates from one
into the other irreducible representation of the Lorentz group.

Let us consider now the symmetries of the two-component complex Majorana
equation, in particular the charge exchange $\mathcal{C}$, parity
$\mathcal{P}$, and time reversal $\mathcal{T}$ operations. Here we expand the
lucid discussion of Case \cite{case}. Generally speaking the Majorana
equation is invariant under the symmetry operation $\mathcal{O}$, if the
two-component spinor
\begin{equation}
\label{eq:12a} \phi^\mathcal{O} = \mathcal{O}\phi
\end{equation}
also fulfils that equation. When applying the operation $\mathcal{O}$ from
the left and its inverse $\mathcal{O}^{-1}$ from the right, whereby the unit
operator is given by the decomposition $\mathcal{O}\mathcal{O}^{-1}=1$, we
obtain the result
\begin{equation}
\label{eq:13a} \left( \mathcal{O}(\mathrm{i}\frac{\partial}{\partial
t})\mathcal{O}^{-1} + \mathcal{O}(\boldsymbol{\sigma} \cdot
\mathrm{i}\frac{\partial}{\partial \mathbf{x}})\mathcal{O}^{-1} - m
\mathcal{O} \tau \mathcal{O}^{-1} \right)\mathcal{O}\phi(\mathbf{x},t)=0.
\end{equation}
We define as usually the time and space coordinate inversion operations
$\mathbb{T}$ and $\mathbb{P}$ on a spinor $\phi$ by
\begin{equation}
\label{eq:14} \mathbb{T} \phi(\mathbf{x},t)= \phi(\mathbf{x},-t),
\end{equation}
\begin{equation}
\label{eq:15} \mathbb{P} \phi(\mathbf{x},t)= \phi(-\mathbf{x},t),
\end{equation}
and also recall the complex conjugation operation $\mathbb{C}$, which gives
$\mathbb{C} \mathrm{i} \mathbb{C}^{-1} = - \mathrm{i}$, and yields
\begin{equation}
\label{eq:16} \mathbb{C} \phi(\mathbf{x},t)= \phi^*(\mathbf{x},t).
\end{equation}
Here the asterisk denotes again the complex conjugate number. With these
preparations in mind, it is easy to see which operators provide the various
symmetry operations, which are composed in the Table~\ref{table:1}. To
complete the operator algebra, it is important to note that the coordinate
reversal operators $\mathbb{T}$ and $\mathbb{P}$ commute with $\tau$ and
$\boldsymbol{\sigma}$, respectively.

\begin{table}
\caption{ {\bf Symmetry operations }} \vspace{0.5 em}
\begin{tabular}{lccc}
\hline
Operation     & Time reversal     &      Parity     &  Chirality (charge)  \\
    &      &           &  conjugation   \\
\hline
Operator & $\mathcal{T}=\tau \mathbb{T}$ &
$\mathcal{P}= \mathrm{i} \tau \mathbb{P}$   & $\mathcal{C} = \tau = \sigma_\mathrm{y}\mathbb{C}$  \\
\hline
\end{tabular}
%      }
%  \end{center}
\label{table:1} %\end{center}
\end{table}

Let us first consider in (\ref{eq:13a}) the time reversal,
$\mathcal{O}=\mathcal{T}=\tau\mathbb{T}$. Apparently, it does not affect the
mass term with $\tau$, and also leaves the first term as well as the kinetic
term $\mathrm{i}\boldsymbol{\sigma}$ unchanged. Therefore,
$\phi^\mathcal{T}=\sigma_\mathrm{y}\phi^*(\mathbf{x},-t)$ solves the Majorana
equation (\ref{eq:10}) as well. Conversely, the parity operation
$\mathcal{O}=\mathcal{P}=\mathrm{i} \tau \mathbb{P}$ anticommutes with the
mass term and inverts the sign of the first term in (\ref{eq:13a}), and it
also does not leave the momentum term invariant, since as $\mathbf{x}$
changes its sign so does the spatial derivative. Consequently,
$\phi^\mathcal{P}=\mathrm{i}\sigma_\mathrm{y}\phi^*(-\mathbf{x},t)$ does
solve the Majorana equation (\ref{eq:10}). Finally, we consider the chirality
conjugation, $\mathcal{O}=\mathcal{C}=\tau$, which only changes the sign of
the first term. Therefore, $\phi^\mathcal{C}= \sigma_\mathrm{y}
\phi^*(\mathbf{x},t)$ does not solve the Majorana equation (\ref{eq:10}) but
solves its conjugate version (\ref{eq:11}). In conclusion, of the symmetry
operations given in Table~\ref{table:1} time reversal and parity are obeyed,
but chirality inversion is maximally broken.

Finally, as the second major topic of our paper we shall reformulate the
two-component Majorana equation in its real form by decomposing the spinor
$\phi$ into its real and imaginary part: $\phi = \phi_R + \mathrm{i} \phi_I$.
Making use of the three real $2\times2$ matrices defined in equation
(\ref{eq:4}) we can write:
\begin{equation}
\label{eq:57a}
\left( \bar{\gamma} \frac{\partial}{\partial t} -  \bar{\alpha} \frac{\partial}{\partial x}
+ \bar{\beta} \frac{\partial}{\partial z} \right) \phi_R +  \frac{\partial}{\partial y} \phi_I = - m \phi_R,
\end{equation}
and similarly we obtain
\begin{equation}
\label{eq:57b}
\left( \bar{\gamma} \frac{\partial}{\partial t} -  \bar{\alpha} \frac{\partial}{\partial x}
+ \bar{\beta} \frac{\partial}{\partial z} \right) \phi_I - \frac{\partial}{\partial y} \phi_R = + m \phi_I,
\end{equation}
both of which can be combined, by introducing $\phi_\pm= \phi_R \pm \phi_I$,
in a single equation:
\begin{equation}
\label{eq:57c}
\left( \bar{\gamma} \frac{\partial}{\partial t} -  \bar{\alpha} \frac{\partial}{\partial x}
+ \bar{\beta} \frac{\partial}{\partial z} \right) \phi_\pm
= \left( \pm \frac{\partial}{\partial y}  -  m \right) \phi_\mp.
\end{equation}
Note that $\bar{\gamma}^2=-1$, $\bar{\alpha}^2=1$, and $\bar{\beta}^2=1$.
Furthermore, these matrices all anticommute, and thus by squaring
(\ref{eq:57c}) one immediately retains the scalar Klein-Gordon equation
reading
\begin{equation}
\label{eq:58}
\left( \frac{\partial^2}{\partial t^2} -  \frac{\partial^2}{\partial \mathbf{x}^2}
 + m^2  \right) \phi_\pm = 0,
\end{equation}
which in fact was our starting point, when deriving a relativistic wave
equation.

From the two real two-component Majorana equations (\ref{eq:57c}), when both
are combined into a single one, the standard Dirac equation in the real
Majorana representation follows immediately. Namely, we may arrange the two
spinors $\phi_\pm$ into a single four-component real Dirac spinor and write
$\psi^\dag= (\phi^\dag_+,\phi^\dag_-)$. Then the coupled system of
(\ref{eq:57c}) transforms into a $4\times4$ real matrix differential equation
which reads
\begin{equation}
\label{eq:59}
\left(
\begin{array}{cc}
\bar{\gamma}  &  0  \\
0  & \bar{\gamma}  \\
\end{array}
\right) \frac{\partial \psi}{\partial t}
- \left(
\begin{array}{cc}
\bar{\alpha}   & 0\\
0 & \bar{\alpha}  \\
\end{array}
\right) \frac{\partial \psi}{\partial x}
+ \left(
\begin{array}{cc}
0  &  -1  \\
1  &  0  \\
\end{array}
\right) \frac{\partial \psi}{\partial y}
+ \left(
\begin{array}{cc}
\bar{\beta}  &  0  \\
0  &   \bar{\beta}  \\
\end{array}
\right) \frac{\partial \psi}{\partial z} = - m
\left(
\begin{array}{cc}
0  &  1  \\
1  &  0  \\
\end{array}
\right)
\psi. \qquad
\end{equation}
Consequently, we may now introduce the subsequent real $4\times4$ Majorana
matrices in their natural (as deduced from the Pauli matrices) representation
\begin{equation}
\label{eq:60}
\bar{\gamma}^\mu=
\left(
\left(
\begin{array}{cc}
0 & \bar{\gamma}  \\
\bar{\gamma} & 0  \\
\end{array}
\right),
\left(
\begin{array}{cc}
0 & -\bar{\alpha}  \\
-\bar{\alpha} &  0 \\
\end{array}
\right),
\left(
\begin{array}{cc}
1  &  0  \\
0  &  -1  \\
\end{array}
\right),
\left(
\begin{array}{cc}
0  & \bar{\beta}  \\
\bar{\beta} &  0  \\
\end{array}
\right)
\right).
\end{equation}
They mutually anticommute and obey: $\bar{\gamma}^\mu\bar{\gamma}^\nu +
\bar{\gamma}^\nu\bar{\gamma}^\mu = - 2g^{\mu\nu}$. Thus the real Dirac
equation in Majorana representation reads
\begin{equation}
\label{eq:61} \bar{\gamma}^\mu \partial_\mu\psi +  m\psi = 0,
\end{equation}
which can, with the help of the purely imaginary Dirac matrices
$\gamma^\mu=\mathrm{i} \bar{\gamma}^\mu$, easily be brought into the standard
form of the Dirac equation:
\begin{equation}
\label{eq:62} \mathrm{i} \gamma^\mu \partial_\mu \psi =  m\psi.
\end{equation}
Therefore, the four-component Dirac equation (\ref{eq:61}) is, in its real
Majorana representation, a direct consequence of the basic two-component
complex Majorana equation (\ref{eq:10}), which was derived here without
invoking the Dirac equation in the first place. Subsequently, we shall also
derive the four-component real eigenspinors of equation (\ref{eq:61}).

\section{Eigenfunctions of the two-component Majorana equation with mass term}
\label{s3}

Returning to the basic equation (\ref{eq:10}), we here derive its
eigenfunctions. In order to solve it, we make the usual plane-wave ansatz,
$\phi(\mathbf{x},t) = u(\mathbf{p},E) \exp(-\mathrm{i} Et + \mathrm{i}
\mathbf{p}\cdot\mathbf{x}) + v(\mathbf{p},E) \exp(\mathrm{i} Et - \mathrm{i}
\mathbf{p}\cdot\mathbf{x})$. But note that we require the plane wave and its
complex conjugate as well, because of the operator $\tau$ in (\ref{eq:10}).
The two-component spinors $u$ and $v$ can not be assumed to be complex
conjugated, i.e. the wave function cannot be expected to be real. The
resulting linked eigenspinor equations are
\begin{equation}
\label{eq:20} (E - {\boldsymbol{\sigma}} \cdot \mathbf{p}) u(\mathbf{p},E)= m
\tau v(\mathbf{p},E),
\end{equation}
\begin{equation}
\label{eq:21} (E - {\boldsymbol{\sigma}} \cdot \mathbf{p}) v(\mathbf{p},E)= -
m \tau u(\mathbf{p},E).
\end{equation}
By insertion of the first into the second equation, or vice versa, the
relativistic dispersion relation is obtained
\begin{equation}
\label{eq:20a} \left( \tau (E - {\boldsymbol{\sigma}} \cdot \mathbf{p}) \tau
(E - {\boldsymbol{\sigma}} \cdot \mathbf{p}) + m^2 \right) u(\mathbf{p},E)= 0
\end{equation}
which yields the two eigenvalues
\begin{equation}
\label{eq:22} E_{1,2}(\mathbf{p}) = \pm \sqrt{m^2 + \mathbf{p}^2},
\end{equation}
which are obtained from the requirement that for nontrivial solutions of the
spinors $u$ and $v$ to exist the determinant associated with the eigenvalue
problem (\ref{eq:20a}) must vanish. The negative root in (\ref{eq:22}) cannot
be neglected, since as usually it is related to antiparticles. We can solve
(\ref{eq:21}) for $v$ and insert it back into the above ansatz for
$\phi(\mathbf{x},t)$ to obtain finally the solutions of (\ref{eq:10}) and
(\ref{eq:11}) in the concise forms
\begin{equation}
\label{eq:23} \phi(\mathbf{x},t) = (1- \frac{E + {\boldsymbol{\sigma}}\cdot
\mathbf{p}}{m} \tau) \exp(-\mathrm{i} Et + \mathrm{i}
\mathbf{p}\cdot\mathbf{x}) u,
\end{equation}
\begin{equation}
\label{eq:24}
\chi(\mathbf{x},t)=\tau\phi(\mathbf{x},t) = (\tau + \frac{E
-{\boldsymbol{\sigma}}\cdot \mathbf{p}}{m}) \exp(-\mathrm{i} Et + \mathrm{i}
\mathbf{p}\cdot\mathbf{x}) u.
\end{equation}
To validate these solutions by direct differentiation, careful attention must
be paid to the anticommutation rules between $\tau$ and $\mathrm{i}$,
respectively $\boldsymbol{\sigma}$. We may also solve (\ref{eq:21}) for $u$
instead, and then it insert it back into the above ansatz for
$\phi(\mathbf{x},t)$.

We are free to choose for the eigenspinor $u$ the two standard spin up and
down eigenfunctions: $u^\dagger_1=(1,0)$ and $u^\dagger_2=(0,1)$, and
similarly for $v$, but there is a better and more adequate choice if $p$ is
nonzero, as discussed below. Similar solutions like that of equations
(\ref{eq:23}) and (\ref{eq:24}) are obtained for the negative energy root in
(\ref{eq:20}), yielding the antiparticle wavefunctions. Superposition of all
the Fourier modes and their summation over the momentum variable $\mathbf{p}$
leads finally to the general Majorana fields, whose quantization then follows
from the canonical rules [2-5] and are given below. Above we obtained the
formal solutions of the Majorana equation, which we reiterate here by
introducing the Majorana operator $\mathcal{M}$ as follows
\begin{equation}
\label{eq:27} \mathcal{M}= \mathrm{i}\left( \frac{\partial}{\partial t} +
\boldsymbol{\sigma} \cdot \frac{\partial}{\partial \mathbf{x}} \right)
- m \tau.
\end{equation}
The solution $\phi$ obeys $\mathcal{M}\phi=0$. Note, however, that the
four-momentum operator $P_\mu= \mathrm{i}\partial_\mu =
\mathrm{i}(\partial/\partial t,\partial/\partial \mathbf{x})$ does not
commute with $\mathcal{M}$, and neither does the spin-flip operator $\tau$.
Therefore, the $\phi$ and $\chi$ of the equations (\ref{eq:23}) to
(\ref{eq:24}) are not eigenfunctions of any of these operators. However, the
helicity operator, $\boldsymbol{\sigma}\cdot \mathbf{P}$, does commute with
$\mathcal{M}$.

Consequently, to chose the eigenfunctions of the helicity operator in $\phi$
for the free functions $u$ or $v$ is most convenient. The eigenvalue equation
of the helicity operator in Fourier space reads
\begin{equation}
\label{eq:28} ({\boldsymbol{\sigma}} \cdot \hat{\mathbf{p}})
u_\pm(\hat{\mathbf{p}})= \pm u_\pm(\hat{\mathbf{p}}).
\end{equation}
The two eigenvectors depend on the momentum unit vector
$\hat{\mathbf{p}}=\mathbf{p}/p$ and are given by
\begin{equation}
\label{eq:29} u_+ (\hat{\mathbf{p}}) = \left(
\begin{array}{c}
\cos \frac{\theta}{2} \; \mathrm{e}^{-\frac{\mathrm{i}}{2} \phi}\\
\sin \frac{\theta}{2} \; \mathrm{e}^{ \frac{\mathrm{i}}{2} \phi}\\
\end{array}
\right), \;\;\; u_- (\hat{\mathbf{p}}) = \left(
\begin{array}{c}
- \sin \frac{\theta}{2} \; \mathrm{e}^{-\frac{\mathrm{i}}{2} \phi} \\
  \cos \frac{\theta}{2} \; \mathrm{e}^{ \frac{\mathrm{i}}{2} \phi} \\
\end{array}
\right),
\end{equation}
in which the half angles of $\theta$ and $\phi$ appear. The eigenvectors for
the same $\hat{\mathbf{p}}$ are orthogonal to each other and normalized to a
modulus of unity, and they obey the relation $u_\pm (-\hat{\mathbf{p}}) =
u_\mp (\hat{\mathbf{p}})$. This is a consequence of the eigenvalue equation
(\ref{eq:28}), which implies that $u_\pm (\hat{\mathbf{p}})$ is an
eigenvector of the helicity operator, corresponding to a right-handed,
respectively left-handed, screw with respect to the momentum direction.
According to (\ref{eq:29}) we have $u^\dag_\pm(\hat{\mathbf{p}})
u_\pm(\hat{\mathbf{p}}) =1$ and $u^\dag_\mp(\hat{\mathbf{p}})
u_\pm(\hat{\mathbf{p}}) =0$. The dagger denotes as usually the transposed
(denoted by the superscript $T$) and complex conjugated vector, respectively
matrix. The scalar product between two vectors (spinors) $v$ and $w$ is just
the sum over the products of their two components, i.e. $v w$ simply stands
for $v_1w_1+v_2w_2$.

We emphasize that the spin-flip operator $\tau = \sigma_y \mathbb{C}$, when
operating on the above eigenspinors, leads to
\begin{equation}
\label{eq:30}
\tau u_\pm(\hat{\mathbf{p}})
= \pm \mathrm{i} u_\mp(\hat{\mathbf{p}}),
\end{equation}
i.e., it connects the eigenfunctions of opposite helicity, and turns out to
be a quite useful relation in the subsequent considerations. By its
application, we can write the two possible associated eigenfunctions after
(\ref{eq:23}) as
\begin{equation}
\label{eq:31}
\phi_\pm(\mathbf{x},t) = \exp(-\mathrm{i} Et + \mathrm{i}
\mathbf{p}\cdot\mathbf{x})u_\pm(\hat{\mathbf{p}})
\mp \frac{\mathrm{i}}{m}(E + {\boldsymbol{\sigma}}\cdot
\mathbf{p}) \exp(\mathrm{i} Et - \mathrm{i}
\mathbf{p}\cdot\mathbf{x}) u_\mp(\hat{\mathbf{p}}),
\end{equation}
where the advantage of the helicity eigenfunctions becomes obvious. Namely,
by use of (\ref{eq:28}) we obtain the (now normalized to unity)
eigenfunction,
\begin{equation}
\label{eq:32}
\phi_\pm(\mathbf{x},t) = \frac{1}{\sqrt{2E}}
( \sqrt{E \pm p}\,\exp(-\mathrm{i} Et + \mathrm{i}
\mathbf{p}\cdot\mathbf{x})u_\pm(\hat{\mathbf{p}})
\mp \mathrm{i} \sqrt{E \mp p}\, \exp(\mathrm{i} Et - \mathrm{i}
\mathbf{p}\cdot\mathbf{x}) u_\mp(\hat{\mathbf{p}}) ), \qquad
\end{equation}
which is a mixed state involving both helicities in a symmetric fashion. Note
that for vanishing mass, $m=0$ corresponding to the Weyl equation, only the
positive helicity remains, and thus the wavefunction becomes purely
right-handed. Its left-handed version is then obtained by applying the
spin-flip operator $\tau$ on (\ref{eq:32}) like in equation (\ref{eq:24}).
Operation of $\mathcal{M}$ on this $\phi_\pm$ validates that it solves the
Majorana equation, i.e. $\mathcal{M}\phi_\pm=0$. A simple form of the
eigenspinor is obtained for a particle at rest, i.e. $p=0$, which yields
\begin{equation}
\label{eq:33}
\phi_\pm(t) = \frac{1}{\sqrt{2}} \left( \exp(-\mathrm{i} m t )u_\pm
\mp \mathrm{i}  \exp(+\mathrm{i} m t ) u_\mp \right).
\end{equation}
The helicity eigenvectors in this case can be chosen to be the standard ones
obtained for the angles $\phi=0$ and $\theta=0$ according to (\ref{eq:29}).

Concerning the adequate choice of the eigenspinor $u_{1,2}$ in (\ref{eq:23})
for the two possible eigenvalues $E_{1,2}(p)=\pm\sqrt{m^2+p^2}$, we recall
that the helicity operator does commute with $\mathcal{M}$, and generally we
should consider a mixture or superposition of both helicities. But this is
indeed already implied in the solution (\ref{eq:32}). Therefore, as we can
only have two linearly independent eigenvectors to define the basis, we
assume that $u$ can be decomposed such that these two eigenvectors are
defined by
\begin{equation}
\label{eq:34a}
u_{1,2}(\mathbf{p}) = a_{1,2}(\mathbf{p}) u_\pm(\hat{\mathbf{p}}),
\end{equation}
with some complex amplitude $a_{1,2}(\mathbf{p})$ of module unity,
$|a_{1,2}|^2=1$, to ensure normalization. The natural but arbitrary
association with the sign of the energy (\ref{eq:22}) is to take the positive
sign for positive (right-handed) helicity and the negative sign for negative
(left-handed) helicity. Thus we may conventionally refer to particles (index
1) and antiparticles (index 2). Upon insertion of this ansatz in
(\ref{eq:23}) we obtain a yet more general expression for the two related
eigenfunctions as follows
\begin{equation}
\label{eq:35a} \phi_1(\mathbf{x},t) =  \sqrt{\frac{E_1 + p}{2E_1}} \,
\exp(-\mathrm{i} E_1 t + \mathrm{i} \mathbf{p}\cdot\mathbf{x}) a_1
u_+  - \mathrm{i} \sqrt{\frac{E_1 - p}{2E_1}}\,
\exp(\mathrm{i} E_1 t - \mathrm{i} \mathbf{p}\cdot\mathbf{x})
a^*_1 u_-,
\end{equation}
\begin{equation}
\label{eq:35b} \phi_2(\mathbf{x},t) =  \sqrt{\frac{E_2 - p}{2E_2}} \,
\exp(-\mathrm{i} E_2 t + \mathrm{i} \mathbf{p}\cdot\mathbf{x}) a_2
u_-  + \mathrm{i} \sqrt{\frac{E_2 + p}{2E_2}}\,
\exp(\mathrm{i} E_2 t - \mathrm{i} \mathbf{p}\cdot\mathbf{x})
a^*_2 u_+,
\end{equation}
where the obvious momentum arguments in $u$ and $a$ have been suppressed for
the sake of lucidity. Apparently, the previous wavefunction (\ref{eq:32}) is
retained for $a_{1,2}(\mathbf{p})=1$, and $E_{1,2}=E$. It turns out to be
convenient to introduce the two real quantities
\begin{equation}
\label{eq:36}
\varepsilon_\pm (p)= \sqrt{\frac{E(p) \pm p}{2E(p)} },
\end{equation}
the squares of which add up to unity, $\varepsilon_1^2 + \varepsilon_2^2 =1$.
A useful property of the epsilons is the obvious relation $(E \pm
p)\varepsilon_\mp = m \varepsilon_\pm$. From now on $E(p) = \sqrt{m^2+p^2}$
is used for the positive root obtained from (\ref{eq:22}), and the argument
$p$ in the epsilons will subsequently be omitted for simplicity. Then we can
write the two eigenfunctions (with a new name) concisely as
\begin{equation}
\label{eq:37a} \tilde{\phi}_1(\mathbf{x},t) = \varepsilon_+ \,
\exp(-\mathrm{i} E t + \mathrm{i} \mathbf{p}\cdot\mathbf{x}) a_1
u_+  - \mathrm{i} \varepsilon_-\,
\exp(+\mathrm{i} E t - \mathrm{i} \mathbf{p}\cdot\mathbf{x})
a^*_1 u_-,
\end{equation}
\begin{equation}
\label{eq:37b} \tilde{\phi}_2(\mathbf{x},t) =  \varepsilon_+ \,
\exp(+\mathrm{i} E t + \mathrm{i} \mathbf{p}\cdot\mathbf{x}) a_2
u_-  + \mathrm{i} \varepsilon_-\,
\exp(-\mathrm{i} E t - \mathrm{i} \mathbf{p}\cdot\mathbf{x})
a^*_2 u_+ \,.
\end{equation}
To make the antiparticle wavefunction having the same standard plane wave
phase like that of the particle, we invert its momentum and note again that
the polarization vectors change according to $u_\pm (-\hat{\mathbf{p}}) =
u_\mp (\hat{\mathbf{p}})$. This inversion is permitted as we will later sum
over all momenta, and thus $\mathbf{p}$ is just a mute index. Consequently,
we take $\phi_\mathrm{P}(\mathbf{p}) = \tilde{\phi}_1(\mathbf{p})$ for the
particle, but $\phi_\mathrm{A}(\mathbf{p}) = \tilde{\phi}_2(-\mathbf{p})$ for
the antiparticle, and we redefine the amplitudes as
$a(\mathbf{p})=a_1(\mathbf{p})$ for the particle, but
$b(\mathbf{p})=\mathrm{i} a_2^*(-\mathbf{p})$, respectively,
$b^*(\mathbf{p})=-\mathrm{i} a_2(-\mathbf{p})$ for the antiparticle. The new
wavefunctions finally read as follows
\begin{equation}
\label{eq:38a} \phi_\mathrm{P}(\mathbf{x},t) = \varepsilon_+ \,
\exp(-\mathrm{i} E t + \mathrm{i} \mathbf{p}\cdot\mathbf{x}) a
u_+  - \mathrm{i} \varepsilon_-\,
\exp(+\mathrm{i} E t - \mathrm{i} \mathbf{p}\cdot\mathbf{x})
a^* u_-,
\end{equation}
\begin{equation}
\label{eq:38b} \phi_\mathrm{A}(\mathbf{x},t) =  \varepsilon_- \,
\exp(-\mathrm{i} E t + \mathrm{i} \mathbf{p}\cdot\mathbf{x}) b
u_-  + \mathrm{i} \varepsilon_+\,
\exp(+\mathrm{i} E t - \mathrm{i} \mathbf{p}\cdot\mathbf{x})
b^* u_+ \,.
\end{equation}

Both eigenfunctions are normalized to unity, yielding the relation
$\phi^\dag_\mathrm{P,A}\phi_\mathrm{P,A}=1$, which is obtained by using
$\varepsilon_+^2+\varepsilon_-^2=1$, exploiting the normalization and
orthogonality of the helicity eigenvectors, and noting that $a^*a=1$ and
$b^*b=1$. They form a complete eigenvector basis of the Majorana operator
$\mathcal{M}$ and obey $\mathcal{M}\phi_\mathrm{P,A}=0$. However, we
emphasize again that they are neither eigenfunctions of the four-momentum
operator $P^\mu= (P_0, \mathbf{P}$), nor of the helicity operator or the
spin-reversal operator $\tau$.

By adding up the the two contributions we retain the most general solution of
the Majorana equation (\ref{eq:10}) as the two-component spinor in the form,
\begin{equation}
\label{eq:38c}
\phi(\mathbf{x},t) = \phi_\mathrm{P}(\mathbf{x},t) + \phi_\mathrm{A}(\mathbf{x},t).
\end{equation}
This superposition of the particle and antiparticle eigenspinors yields the
full Majorana wavefunction in the form
\begin{equation}
\label{eq:40}
\phi(\mathbf{x},t) = \exp(-\mathrm{i} E t + \mathrm{i} \mathbf{p}\cdot\mathbf{x})
(\varepsilon_+  u_+ a + \varepsilon_- u_- b) - \mathrm{i} \exp(+\mathrm{i} E t - \mathrm{i}
\mathbf{p}\cdot\mathbf{x})( \varepsilon_- u_- a^* - \varepsilon_+ u_+ b^*).
\end{equation}

\section{The Majorana quantum field}
\label{s4}

At this point the transition from (\ref{eq:40}) to a quantum field is quite
natural and obvious. We just have to replace the amplitudes in (\ref{eq:40})
by the canonical anti-commuting fermion operators obeying
\begin{equation}
\label{eq:39a} \{ a(\mathbf{p}), a^\dag (\mathbf{p}^\prime) \} = \delta_{\mathbf{p}, \mathbf{p}^\prime},
\end{equation}
\begin{equation}
\label{eq:39b} \{ b(\mathbf{p}), b^\dag (\mathbf{p}^\prime) \} = \delta_{\mathbf{p}, \mathbf{p}^\prime}.
\end{equation}
Of course, all possible anticommutators between either a pair of creation,
respectively destruction, operators are zero. Mutually, between the $a$s and
$b$s the anticommutators vanish, as they should since these two degrees of
freedom are independent. The operator $a^\dag (\mathbf{p})$ creates, and vice
versa $a(\mathbf{p})$ annihilates the plane-wave state of a particle of
positive helicity, with momentum $\mathbf{p}$ and energy $E=\sqrt{m^2+p^2}$.
The $b$-operators do the same, yet for the related antiparticle of the
opposite negative helicity.

We finally get, by keeping explicitly all momentum arguments, the following
field operator for any given momentum $\mathbf{p}$ in the concise form
\begin{equation}
\label{eq:44}
\Phi_\mathbf{p}(\mathbf{x},t) = \exp(-\mathrm{i} E(p) t + \mathrm{i}
\mathbf{p}\cdot\mathbf{x}) \underline{c}(\mathbf{p})
- \mathrm{i} \exp(+\mathrm{i} E(p)t - \mathrm{i} \mathbf{p}\cdot\mathbf{x})
\underline{d}^\dag(\mathbf{p}),
\end{equation}
which assumes the standard form (similar to what is known from the Dirac
equation). The quantum field operators corresponding to $\phi$ are denoted by
a capital $\Phi$. We can now sum up over all momenta to obtain the full
Majorana quantum field
\begin{equation}
\label{eq:43}
\Phi_\mathcal{M}(\mathbf{x},t) = \sum_\mathbf{p} \Phi_\mathbf{p}(\mathbf{x},t).
\end{equation}

Above we introduced above as abbreviations two operators, which correspond to
the creation, respectively annihilation, polarization-vector operators
(indicated by an underscore) for a given mixed helicity state as follows:
\begin{equation}
\label{eq:45}
\underline{c}(\mathbf{p}) = \varepsilon_+(p) u_+(\hat{\mathbf{p}}) a(\mathbf{p}) +
\varepsilon_-(p) u_-(\hat{\mathbf{p}})  b(\mathbf{p}),
\end{equation}
and similarly
\begin{equation}
\label{eq:46}
\underline{d}^\dag(\mathbf{p}) = \varepsilon_-(p) u_-(\hat{\mathbf{p}}) a^\dag(\mathbf{p})
- \varepsilon_+(p) u_+(\hat{\mathbf{p}}) b^\dag(\mathbf{p}).
\end{equation}
They obey the standard anticommutation rule, whereby the inner product
between the original helicity eigen-vector is used, as well as the relation
$\varepsilon_+^2 + \varepsilon_-^2 =1$. Thus we obtain, by implying the
standard scalar product between the complex two-component spinors involved,
the result
\begin{equation}
\label{eq:47} \{ \underline{c}(\mathbf{p}), \underline{c}^\dag (\mathbf{p}^\prime) \}
= \delta_{\mathbf{p}, \mathbf{p}^\prime},
\end{equation}
and correspondingly
\begin{equation}
\label{eq:48} \{ \underline{d}(\mathbf{p}), \underline{d}^\dag (\mathbf{p}^\prime) \} =
\delta_{\mathbf{p}, \mathbf{p}^\prime}.
\end{equation}
Again all possible mixed anticommutators between the $\underline{c}$ and
$\underline{d}$ operators vanish. For the two particle number operators we
obtain from the above equations (\ref{eq:45}) and (\ref{eq:46}) the linear
combinations
\begin{equation}
\label{eq:51a}
\underline{c}^\dag \underline{c} =
 \varepsilon_+^2 a^\dag a + \varepsilon_-^2  b^\dag b,
\end{equation}
\begin{equation}
\label{eq:51b}
\underline{d}^\dag \underline{d} =
 \varepsilon_-^2 a^\dag a + \varepsilon_+^2  b^\dag b.
\end{equation}

This combined coordinate transformation, involving the original eigenspinors
of the helicity operator and their corresponding state operators (classically
amplitudes), has a physically important interpretation. As stressed before,
the helicity operator does commute with $\mathcal{M}$, but one requires a
mixed helicity state to represent its eigenfunctions. This mixing depends on
the ratio of the momentum, $p$, to the total energy, $E(p)$, of a particle.
Only if $m=0$, we have pure left- or right-handed states, otherwise the
massive states are mixed. That $\mathcal{M}\Phi_\mathbf{p}=0$, follows
readily from the notion that
\begin{equation}
\label{eq:45a}  (E - {\boldsymbol{\sigma}} \cdot \mathbf{p})\underline{c}(\mathbf{p})
= m( \varepsilon_-(p) u_+(\hat{\mathbf{p}}) a(\mathbf{p}) +
\varepsilon_+(p) u_-(\hat{\mathbf{p}})  b(\mathbf{p}) )
= + m\mathrm{i} \tau \underline{d}^\dag(\mathbf{p}),
\end{equation}
respectively, that
\begin{equation}
\label{eq:45b}  (E - {\boldsymbol{\sigma}} \cdot \mathbf{p})\underline{d}^\dag(\mathbf{p})
= m( \varepsilon_+(p) u_-(\hat{\mathbf{p}}) a^\dag(\mathbf{p}) -
\varepsilon_-(p) u_+(\hat{\mathbf{p}})  b^\dag(\mathbf{p}) )
= - m\mathrm{i} \tau \underline{c}(\mathbf{p}),
\end{equation}
where the auxiliary relation $(E \pm p)\varepsilon_\mp = m \varepsilon_\pm$
has been used. We should mention here, that according to its definition the
operation of $\tau$ on a creation or annihilation operator is defined such
that $\tau a =a^\dag \tau$, or similarly $\tau a^\dag = a \tau$.

As an interlude, we may now consider the unmixed massless case $m=0$, which
gives us the Weyl field. Then $\varepsilon_+=1$ and $\varepsilon_-=0$. The
resulting quantum field operator is given by
\begin{equation}
\label{eq:49a}
\Phi_{\mathcal{W}\,\mathrm{R}}(\mathbf{x},t) = \sum_\mathbf{p}u_+(\hat{\mathbf{p}})(\exp(-\mathrm{i} p t + \mathrm{i} \mathbf{p}\cdot\mathbf{x}) a(\mathbf{p})
 +  \mathrm{i} \exp(+\mathrm{i} p t - \mathrm{i}\mathbf{p}\cdot\mathbf{x}) b^\dag(\mathbf{p}) ).
\end{equation}
Apparently, this field involves only the polarization vector for positive
helicity. Therefore, in this case the right-chiral Weyl field operator
annihilates right-handed particles and creates left-handed antiparticles.
Operating with the spin flip operator $\tau$, according to $\tau a \tau^{-1}=
a^\dag$ and equation (\ref{eq:30}), on this field yields the left-chiral Weyl
field, which reads
\begin{equation}
\label{eq:49b}
\Phi_{\mathcal{W}\,\mathrm{L}}(\mathbf{x},t) = \sum_\mathbf{p}u_-(\hat{\mathbf{p}})(\exp(-\mathrm{i} p t + \mathrm{i} \mathbf{p}\cdot\mathbf{x})
b(\mathbf{p})  +  \mathrm{i} \exp(+\mathrm{i} p t - \mathrm{i}
\mathbf{p}\cdot\mathbf{x}) a^\dag(\mathbf{p}) ).
\end{equation}
Consequently, in the case of the left-chiral Weyl field the operator creates
right-handed particles and annihilates left-handed antiparticles.

Similarly, we obtain for the massive right-chiral Majorana quantum field
\begin{equation}
\label{eq:55a}
\Phi_{\mathcal{M}\,\mathrm{R}}(\mathbf{x},t) = \sum_\mathbf{p} ( \exp(-\mathrm{i} E(p) t + \mathrm{i}
\mathbf{p}\cdot\mathbf{x}) \underline{c}(\mathbf{p}) - \mathrm{i} \exp(+\mathrm{i} E(p)
t - \mathrm{i} \mathbf{p}\cdot\mathbf{x}) \underline{d}^\dag(\mathbf{p}),
\end{equation}
and by application of $\tau$ on it the left-chiral Majorana quantum field
\begin{equation}
\label{eq:55b}
\Phi_{\mathcal{M}\,\mathrm{L}}(\mathbf{x},t) = \sum_\mathbf{p} ( \exp(-\mathrm{i} E(p) t + \mathrm{i}
\mathbf{p}\cdot\mathbf{x}) (\mathrm{i}\tau \underline{d}^\dag(\mathbf{p})) - \mathrm{i} \exp(+\mathrm{i} E(p)
t - \mathrm{i} \mathbf{p}\cdot\mathbf{x}) (\mathrm{i}\tau \underline{c}(\mathbf{p})).
\end{equation}
The corresponding operators can be read off equations (\ref{eq:45a}) and
(\ref{eq:45b}).

Ultimately, we are interested in the expectation value of a given hermitian
operator $\mathcal{O}$ for the quantum field $\Phi$, for example the Majorana
fields (\ref{eq:55a}) and (\ref{eq:55b}) or the Weyl fields of (\ref{eq:49a})
and (\ref{eq:49b}), which are all not hermitian fields. For the massless Weyl
fields, the related plane waves are also eigenfunctions of the four-momentum
operator $P^\mu$. In contrast, for the massive Majorana field this is not
true, and therefore only the expectation value of $P^\mu$ can be evaluated.
The same comment applies to the kinetic helicity operator, which yet does
commute with $\mathcal{M}$. This interesting feature was discussed at length
and clarified by Mannheim \cite{mannheim}, who concluded that for the massive
Majorana field theory ``second quantization is necessary a priori in order to
produce a sensible particle interpretation''. Therefore, we can only
calculate average expectation values, which are defined by volume integrals
of the real part of the binary form
\begin{equation}
\label{eq:41}
O = <\mathcal{O}> = \frac{1}{2} \int d^3x \left( \Phi^\dag \mathcal{O}
\Phi + (\mathcal{O} \Phi)^\dag \Phi \right).
\end{equation}
This formula shall be now applied to calculate the expectation value of the
energy-momentum four-vector, the helicity and number operator of the
right-chiral Majorana quantum field
$\Phi_{\mathcal{M}\,\mathrm{R}}(\mathbf{x},t)$ of equation (\ref{eq:55a}).
Upon its insertion into (\ref{eq:41}), and after some lengthy but standard
calculations (similar to what is usually done with the Dirac equation, e.g.
see the textbooks \cite{kaku} or \cite{das}) we obtain the concise result
\begin{equation}
\label{eq:42}
P^\mu_{\mathcal{M}\,\mathrm{R}} =  <(P_0, \mathbf{P})>_{\mathcal{M}\,\mathrm{R}}  =  \sum_\mathbf{p}
\left(\sqrt{m^2+p^2}, \mathbf{p}\right)\left(a^\dag(\mathbf{p}) a(\mathbf{p})
+ b^\dag(\mathbf{p}) b(\mathbf{p})\right),
\end{equation}
where conventionally an infinite constant, $-\sum_\mathbf{p}E(p)$, has been
discarded as irrelevant zero-point energy. Let us consider the helicity
density at a given $\mathbf{p}$ of the Majorana operator (\ref{eq:44}). We
find
\begin{equation}
\label{eq:50}
\Phi^\dag_\mathbf{p} ({\boldsymbol{\sigma}} \cdot \hat{\mathbf{p}}) \Phi_\mathbf{p} =
\underline{c}^\dag(\mathbf{p}) ({\boldsymbol{\sigma}} \cdot \hat{\mathbf{p}}) \underline{c}(\mathbf{p}) +
\underline{d}(\mathbf{p}) ({\boldsymbol{\sigma}} \cdot \hat{\mathbf{p}}) \underline{d}^\dag(\mathbf{p})
= a^\dag a - b^\dag b - \varepsilon_-^2 + \varepsilon_+^2. \qquad
\end{equation}
and thus by using these relations and by summing up all contributions of
equation (\ref{eq:50}), we obtain the total helicity operator
$H_{\mathcal{M}\,\mathrm{R}}$, which is (within an unimportant constant)
given by
\begin{equation}
\label{eq:51}
H_{\mathcal{M}\,\mathrm{R}} = < \boldsymbol{\sigma} \cdot \hat{\mathbf{P}} >_{\mathcal{M}\,\mathrm{R}}\,
 = \sum_\mathbf{p} \left( a^\dag(\mathbf{p}) a(\mathbf{p})
- b^\dag(\mathbf{p}) b(\mathbf{p})\right).
\end{equation}
It involves the difference of the number operators of the particles and
antiparticles, which therefore are expected to contribute oppositely to the
net helicity.

Interestingly, we can go through the same procedure with the left-chiral
field (\ref{eq:55b}) and end up with exactly the same expectation values as
calculated above. One gets a similar result for the Weyl quantum field, by
putting either $m=0$ in the Majorana field or by using directly equations
(\ref{eq:49a}) and (\ref{eq:49b}). In other words it suffices to consider
mathematically the original and constitutive equation (\ref{eq:7}). Chirality
conjugation, i.e. the replacement of the spin $\boldsymbol{\sigma}$ by its
negative inverted vector $-\boldsymbol{\sigma}$, does not yield new physical
information concerning the above expectation values (\ref{eq:42}) and
(\ref{eq:51}).

\section{Four-component eigenspinors of the Majorana equation}
\label{s5}

Following the derivation of the real Majorana equation (\ref{eq:61}), its
eigenfunctions can, by construction according to equation (\ref{eq:59}),
easily be derived. Starting point is the complex two-component original wave
function (\ref{eq:40}), in which the wave amplitudes $a$ and $b$ correspond
after the previous section to particles, respectively antiparticles, and thus
transform into the related annihilation operators for the quantum fields. The
four-component eigenfunction is therefore given by the spinor
\begin{equation}
\label{eq:554}
\psi = \left(
\begin{array}{c}
\phi_\mathrm{R}  + \phi_\mathrm{I} \\
\phi_\mathrm{R}  - \phi_\mathrm{I} \\
\end{array}
\right),
\end{equation}
where the real and imaginary parts are taken from (\ref{eq:40}). This spinor
can be written as a sum of the particle and antiparticle components and reads
\begin{equation}
\label{eq:555} \psi(\mathbf{x},t) = \psi_\mathrm{P}(\mathbf{x},t) +
\psi_\mathrm{A}(\mathbf{x},t).
\end{equation}
These two contributions can be expressed, in terms of the complex
four-component polarization spinors to be defined below, separately as
follows:
\begin{eqnarray}
\label{eq:556}
\psi_\mathrm{P}(\mathbf{x},t) = \exp(-\mathrm{i} E t + \mathrm{i} \mathbf{p}\cdot\mathbf{x})
a(\mathbf{p}) \tilde{\alpha}(\mathbf{p}) + \exp(+\mathrm{i} E t - \mathrm{i} \mathbf{p}\cdot\mathbf{x})
a^*(\mathbf{p}) \tilde{\alpha}^*(\mathbf{p}),
\end{eqnarray}
\begin{eqnarray}
\label{eq:557}
\psi_\mathrm{A}(\mathbf{x},t) = \exp(-\mathrm{i} E t + \mathrm{i} \mathbf{p}\cdot\mathbf{x})
b(\mathbf{p}) \tilde{\beta}(\mathbf{p}) + \exp(+\mathrm{i} E t - \mathrm{i} \mathbf{p}\cdot\mathbf{x})
b^*(\mathbf{p}) \tilde{\beta}^*(\mathbf{p}).
\end{eqnarray}
Apparently, $\psi_\mathrm{P,A}=\psi^*_\mathrm{P,A}$, and thus the wave
functions are real. The polarization spinors $\tilde{\alpha}$ and $\tilde{\beta}$ can be
constructed from the two-component eigenfunctions, given in equations
(\ref{eq:29}), of the helicity operator (\ref{eq:28}). Thus they can be
concisely written as
\begin{equation}
\label{eq:558}
\tilde{\alpha}(\mathbf{p}) = \frac{1}{2}\left(
\begin{array}{c}
(\varepsilon_+(p) u_+(\hat{\mathbf{p}}) - \varepsilon_-(p) u^*_-(\hat{\mathbf{p}}))(1-\mathrm{i}) \\
(\varepsilon_+(p) u_+(\hat{\mathbf{p}}) + \varepsilon_-(p) u^*_-(\hat{\mathbf{p}}))(1+\mathrm{i}) \\
\end{array}
\right),
\end{equation}
\begin{equation}
\label{eq:559}
\tilde{\beta}(\mathbf{p}) = \frac{1}{2}\left(
\begin{array}{c}
(\varepsilon_-(p) u_-(\hat{\mathbf{p}}) + \varepsilon_+(p) u^*_+(\hat{\mathbf{p}}))(1-\mathrm{i}) \\
(\varepsilon_-(p) u_-(\hat{\mathbf{p}}) - \varepsilon_+(p) u^*_+(\hat{\mathbf{p}}))(1+\mathrm{i}) \\
\end{array}
\right).
\end{equation}
Use has again been made of the symbols $\varepsilon_\pm$ as defined in
equation (\ref{eq:36}). Using their properties and the orthogonality of
$u_\pm$, one can readily show that $\tilde{\alpha}^\dag \tilde{\alpha}=1$ and $\tilde{\beta}^\dag
\tilde{\beta}=1$, respectively, $\tilde{\alpha}^\dag \tilde{\beta}=0=\tilde{\beta}^\dag \tilde{\alpha}$.

We may now insert the spinors (\ref{eq:556}) and (\ref{eq:557}) into the real
Majorana equation (\ref{eq:61}) or complex Dirac equation (\ref{eq:62}), to
validate that they are solutions of those equations. With the covariant
four-momentum $p_\mu=(E,-\mathbf{p})$, one finds that the real and imaginary
parts of $\tilde{\alpha}=\tilde{\alpha}_\mathrm{R} +
\mathrm{i}\tilde{\alpha}_\mathrm{I}$ (respectively $\tilde{\beta}$) must obey
the coupled equations:
\begin{equation}
\label{eq:560} \bar{\gamma}^\mu p_\mu \tilde{\alpha}_\mathrm{I} +  m\tilde{\alpha}_\mathrm{R} = 0,
\end{equation}
and similarly
\begin{equation}
\label{eq:561} \bar{\gamma}^\mu p_\mu \tilde{\alpha}_\mathrm{R} -  m\tilde{\alpha}_\mathrm{I} = 0,
\end{equation}
which can be combined to yield the complex solution of the Dirac equation as
\begin{equation}
\label{eq:562} \gamma^\mu p_\mu \tilde{\alpha} =  m \tilde{\alpha}.
\end{equation}
The same equation must hold true for the polarization spinor $\tilde{\beta}$, and
this is indeed the case. The latter equation can explicitly be written in
matrix form as $\mathbb{M}\tilde{\alpha}=0$, with the non-hermitian matrix
\begin{equation}
\label{eq:570}
\mathbb{M} = \left(
\begin{array}{cccc}
- p_\mathrm{y} + \mathrm{i} m  &  0  & p_\mathrm{x} &  -E - p_\mathrm{z}\\
0  & - p_\mathrm{y} + \mathrm{i} m    & E - p_\mathrm{z} & -p_\mathrm{x}\\
p_\mathrm{x} & -E - p_\mathrm{z}  &  p_\mathrm{y} + \mathrm{i} m  & 0 \\
E - p_\mathrm{z} &  -p_\mathrm{x} & 0 & p_\mathrm{y} + \mathrm{i} m  \\
\end{array}
\right),
\end{equation}
which is the $4\times4$-matrix analogue of the Majorana operator
$\mathcal{M}$ in equation (\ref{eq:27}). The validation of (\ref{eq:562})
requires some lengthy algebraic calculations that shall not be done here.
Nontrivial solutions of $\tilde{\alpha}$ to exist requires that the
determinant of $\mathbb{M}$ vanishes. The corresponding polynomial in the
real variable $E$ is of fourth order, yet only yields the two real roots
already quoted in equation (\ref{eq:22}), corresponding to particles and
antiparticles.

For the sake of completeness we finally give the full four-component
polarization spinor $\alpha$, which in terms of the angles of the momentum
unit vector $\hat{\mathbf{p}}= (\sin\theta \sin\phi, \sin\theta \cos\phi,
\cos\theta)$ and $p$ reads as follows:
\begin{equation}
\label{eq:580}
\tilde{\alpha}(\mathbf{p}) = \frac{1}{2} \left(
\begin{array}{c}
(\varepsilon_+(p)  \cos \frac{\theta}{2} \; \mathrm{e}^{-\frac{\mathrm{i}}{2} \phi} + \varepsilon_-(p) \sin \frac{\theta}{2} \; \mathrm{e}^{+\frac{\mathrm{i}}{2} \phi})(1-\mathrm{i}) \\
(\varepsilon_+(p)  \sin \frac{\theta}{2} \; \mathrm{e}^{+\frac{\mathrm{i}}{2} \phi} - \varepsilon_-(p) \cos \frac{\theta}{2} \; \mathrm{e}^{-\frac{\mathrm{i}}{2} \phi})(1-\mathrm{i}) \\
(\varepsilon_+(p)  \cos \frac{\theta}{2} \; \mathrm{e}^{-\frac{\mathrm{i}}{2} \phi} - \varepsilon_-(p) \sin \frac{\theta}{2} \; \mathrm{e}^{+\frac{\mathrm{i}}{2} \phi})(1+\mathrm{i}) \\
(\varepsilon_+(p)  \sin \frac{\theta}{2} \; \mathrm{e}^{+\frac{\mathrm{i}}{2} \phi} + \varepsilon_-(p) \cos \frac{\theta}{2} \; \mathrm{e}^{-\frac{\mathrm{i}}{2} \phi})(1+\mathrm{i}) \\
\end{array}
\right).
\end{equation}
Similarly, the polarization spinor $\tilde{\beta}$ can be derived explicitly
from (\ref{eq:559}). This derivation essentially concludes the section on the
solution of the real Majorana equation with mass term. We have presented its
eigenspinor for particles $\psi_\mathrm{P}$ in equation (\ref{eq:556}) and
for antiparticles $\psi_\mathrm{A}$ in (\ref{eq:557}). The related quantum
fields are readily obtained by interpreting $a$ as annihilation operator, and
by replacing $a^*$ by $a^\dag$ and interpreting it as the particle creation
operator, both together obeying the usual fermion anticommutation rule
(\ref{eq:39a}). The resulting field operator $\Psi_\mathrm{P}$ is not
hermitian, though, as $\tilde{\alpha}^\dag$ is not equal to
$\tilde{\alpha}^*$. The same comments apply to the antiparticle quantum field
described by the creation operator $b^\dag$ and polarization spinor
$\tilde{\beta}^*$.

\section{Summary and conclusions}
\label{s6}

The principal goal of this paper was to rederive and discuss the
two-component and four-component Majorana equations completely on their own
rather than as a spinoff of the chiral Dirac equation. We obtained these
equations including a mass term in a new way by a direct linearization of the
relativistic dispersion relation of a massive particle. Thereby we only made
use of the complex conjugation operator and the Pauli spin matrices,
corresponding to the irreducible representation of the Lorentz group. We then
calculated the eigenfunctions of the complex two-component Majorana equation
and its related quantum field in a transparent way, exploiting the spin-flip
or related chirality conjugation operator. Subsequently, we showed the
four-component Dirac equation in its real Majorana representation to be the
natural outcome of the genuine, two-component and complex Majorana equation
(\ref{eq:10}).

As the analysis of this version of the Majorana equation clearly reveals, the
two associated Majorana particles (with creation operator $a^\dag$ and
similarly $b^\dag$) represent linearly independent states and are not their
own antiparticles, when being defined like in reference \cite{mannheim} in
the sense of having positive or negative energies (\ref{eq:22}). We recall
here that the original eigenfunctions (\ref{eq:40}) and (\ref{eq:580}) result
from a superposition of different states, and thus do neither describe a pure
helicity state nor do they have a well defined four-momentum. These distinct
features only emerge by evaluating the expectation values (\ref{eq:42}) and
(\ref{eq:51}) of the quantum fields. They do in fact describe particles and
antiparticles with opposite helicities, consistently with the standard
perception of neutrinos and antineutrinos.

The symmetries of the two-component Majorana equation were also analysed. It
was shown to obey time reversal and parity, yet apparently violates chirality
conjugation by construction. As the Majorana quantum field is uncharged, the
common term ``charge conjugation" does not appear adequate here, but better
Lorentz-covariant complex conjugation \cite{pal}. Even more appropriately,
while stressing clearly the physical meaning, one may speak of ``chirality
conjugation", which refers to the two conjugate versions of the Majorana
equation having opposite chirality. So, the Majorana equation (\ref{eq:10}),
while breaking chiral symmetry by definition, is as basic as Dirac's
equation, and when considered as a quantum field can describe massive and
uncharged right-handed or left-handed particles or antiparticles,
corresponding to massive neutrinos. In our opinion, the approach starting
from scratch with equation (\ref{eq:7}) gives us new insights into the nature
of the Majorana equation, and shows that it deserves to be considered in its
own right.

\end{document}